\documentclass[prl,twocolumn,epsf,showpacs]{revtex4}
\usepackage{graphicx}
\usepackage{dcolumn}
\usepackage{bm}
\usepackage{epsfig}

\begin{document}
\bibliographystyle{../biblio/prsty}

\title{A one-channel conductor in an ohmic environment:
mapping to a TLL and full counting statistics}

\author{I. Safi$^{1}$ and H.~Saleur$^{2,3}$}

\affiliation{$^1$ Laboratoire de Physique des Solides, Universit\'e
Paris-Sud, 91405 Orsay, France\\${}^2$SPHT CEN Saclay, Gif Sur Yvette 91191, France\\
${}^3$ Department of Physics,
University of Southern California, Los Angeles, CA 90089-0484,USA}

\date{\today}

\begin{abstract}
It is shown that a one-channel coherent conductor in an ohmic
environment can be mapped to the problem of a backscattering
impurity in a Tomonaga-Luttinger liquid (TLL). This allows to
determine non perturbatively the effect of the environment on
$I-V$ curves, and to find an exact relationship between dynamic
Coulomb blockade and shot noise. We investigate critically how
this relationship compares to recent proposals in the literature.
The full counting statistics is determined at zero temperature.

\end{abstract}


\maketitle
\begin{figure}[bh]
\includegraphics[scale=0.4,angle=-90]{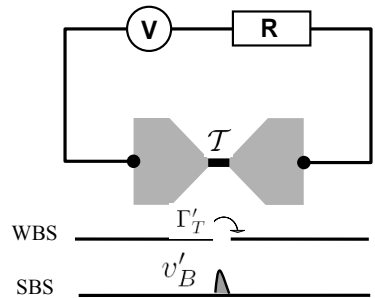}
\caption{\label{fig1}A  quantum circuit of a one-channel coherent
conductor with transmission coefficient ${{\cal T}}$ in series
 with an impedance $Z(\omega)=R$ for $\omega<\omega_R=1/RC$.
 It is mapped to a TLL with parameter $K'=1/(1+r)$ where $r=e^2 R/h$.
  The strong (weak) backscattering limit corresponds to
 the tunnelling (weak backscattering) regime with a dimensionless amplitude $\Gamma'_T$
 ($v'_B$).}
 \end{figure}

 A mesoscopic conductor embedded in an electrical
circuit forms a quantum system violating Ohm's laws. The
transmission/reflection processes of electrons through the
conductor excite the electromagnetic modes of the environment,
 rendering the scattering inelastic, and reducing the
current at low voltage, an effect called environmental Coulomb
blockade\cite{ingold_nazarov}. This picture, valid in the limit of
a weak conductance, changes in the opposite limit of a good
conductance\cite{joyez_multi}. The description of tunnelling via
discrete charge states becomes then ill defined, raising the
question of whether dynamic Coulomb blockade (DCB) survives or is
completely washed out by quantum fluctuations. It is quite clear
that DCB vanishes for a perfectly transmitting conductor. This
property is shared by shot noise which results as well from the
random current pulses due to tunneling events. Such similarity was
concretized \cite{zaikin_noise,levy_yeyati} through a
 challenging relationship between the DCB reduction of the current in a one-channel
 conductor in series with a weak impedance
 and the noise without impedance (see Fig.(\ref{fig1})).
More generally, the DCB variation of the n-1th cumulant of the
current was related to the n-th cumulant without
environment\cite{nazarov_third_cumulant,zaikin}. The environmental
effect on the third cumulant has been the subject of a recent
intensive experimental and theoretical
activity\cite{reulet,nazarov_third_cumulant}.

An ohmic environment could as well simulate the electronic
interactions in the coherent conductor\cite{nazarov_sov}. In this
view, one can wonder whether a one channel conductor in series
with a resistance is equivalent to a one dimensional interacting
system, described by the TLL model\cite{haldane_bosonisation}.
This is already suggested by the power law behavior at small
transmission with an exponent determined by $r=e^2R /h$, the
dimensionless environmental resistance, instead of the microscopic
interaction parameter\cite{ingold_nazarov}.
 Furthermore, Kindermann and Nazarov \cite{nazarov_small_z} have shown recently that at low enough energy, a many-channel
conductor in series with a weak resistance $r\ll 1$ behaves as a
one-channel conductor with an effective energy dependent
transmission ${{\cal T}}(E,r)$ similar to that obtained in a
weakly interacting one-dimensional wire in the presence of a
backscattering center \cite{glazman}. In this framework, the
variation of the current due to DCB is rather given by the shot
noise computed through ${{\cal T}}(E,r)$ instead of the bare
transmission ${{\cal T}}$.

 In this Letter we fully extend
the analogy to a TLL in order to explore the case of an arbitrary
resistance $r$ in series with a coherent one channel conductor
with good transmission. By performing a careful integration over
the
 environmental degrees of freedom, it is
shown that there is an entire low energy regime where the
conductor embedded in its ohmic environment behaves exactly like a
point scatterer in a TLL liquid\cite{remarkingold}, with an
effective parameter $K'=1/(1+r)$.

 In contrast to a renormalization of
the TLL exponent
  by (for instance)
electron-phonon coupling\cite{chen_phonons}, which tend to make
e-e interactions more attractive, here the resistance will rather
induce repulsive interactions, which corresponds to $K'<1$ in the
TLL model. This allows to use exact field theory results obtained
in the TLL context to propose a novel relationship between the DCB
current and shot noise, and more generally between all cumulants.

 Consider first a coherent one-channel conductor coupled to its environment, and
 described by the
total hamiltonian, restricting for simplicity to spinless
electrons:
\begin{equation}\label{Htotal}
H=H_1+H_2+H_{env}+\Gamma_T\psi_1^{\dagger}\psi_2e^{-i\left[e\frac{Vt}{\hbar}-\varphi(t)\right]}+h.c.
\end{equation} Here $H_{1,2}$ is
the electronic Hamiltonian for the right and left electrodes,
$H_{env}$ is a quadratic Hamiltonian describing the
electromagnetic modes of the environment and including the
capacitance of the junction,
%
%
and $V$ is the potential imposed by the voltage generator. The
last term couples the phase across the environmental impedance
$Z(\omega)$ to
 the local electronic fields
$\Psi_{1,2}(0)$ at the end points of the left/right electrodes,
 the momentum dispersion of the tunneling amplitude
$\Gamma_T$ being ignored. In the following, we consider an ideal
Ohmic resistor $R=hr/e^2$. Thus one is restricted to energies
below $\omega_R=1/RC$ since the capacitance of the conductor $C$
is included in the total impedance
$Z(\omega)=R/(1+i\omega/\omega_R)\simeq R$ for $\omega<\omega_R$.
 At zero temperature, the large time
behavior of the phase fluctuations becomes\cite{ingold_nazarov}
\begin{equation}\label{J}
J(t)=\langle\varphi(t)\varphi(0)\rangle-\langle\varphi^2\rangle=-{2r}\ln(i\omega_R
t).
\end{equation}

 The differential dimensionless conductance
 has been computed to lowest order in the
tunnelling amplitude\cite{ingold_nazarov}
\begin{equation}\label{perturbativeI}
G_1=\frac{h}{e^2}\frac{dI}{dV}\approx\frac 1{\Gamma\left( 2
r+1\right)}\left(\frac{\Gamma_T}{h v_F}
\right)^2\left[{e|V|\over\hbar\omega_{R}}\right]^{2r},
\end{equation}
where $\Gamma$ is the gamma function. The similarity with the
power-law behaviors familiar in TLL is striking. Now it will be
shown that it is more than a coincidence.

Since tunnelling is punctual one can use bosonisation for the
electronic part: performing a spherical wave decomposition of the
modes in the electrodes, tunnelling affects only the s waves,
whose dynamics is the same as that of
 one dimensional leads. One introduces the bosonic field $\theta$
 with respect to which the electronic Hamiltonian $H_1+H_2=H_0^{el}$ in Eq.(\ref{Htotal})
 is quadratic, thus its propagator is similar
 to Eq.(\ref{J}):
 \begin{equation}\label{theta}
\langle \theta(t)\theta(0)\rangle=-\ln (i\omega_{F}t).
 \end{equation}

  The tunneling term
  becomes $\psi^{\dagger}_{1}(0)\psi_{2}(0)=
 e^{2i\theta}/{2\pi a}$, with $a$ a distance cutoff, thus the total Hamiltonian Eq.(\ref{Htotal}) reads $H=H_0^{el}+H_{env}+ \frac{\Gamma_T}{2\pi a}
e^{-2i\theta}e^{i\left(\frac{eVt}{\hbar}-\varphi\right)}+h.c.$
Since $\theta$ and $\varphi$ commute, and both $H_0^{el}$ and
$H_{env}$ are quadratic respectively with respect to $\theta$ and
$\varphi$, their sum $H_0'$ is quadratic with respect to the
auxiliary field $\theta'=\theta+\varphi/2$, and $
H=H_{0}'+\frac{\Gamma'_T}{2\pi
a}e^{-2i\theta'(t)-i\frac{eVt}{\hbar}}+h.c.. $  Using
Eqs.(\ref{J},\ref{theta}), one gets $\langle
\theta'(t)\theta'(0)\rangle=-{1\over {2K'}}\ln (i\omega'_F t) $ up
to a constant absorbed in $\Gamma'_T$, the effective tunneling
amplitude. The auxiliary parameter obeys ${1\over K'}\equiv 1+{r}$
and the effective cutoff is $\omega'_F=\min(\omega_R,\omega_F)$.
The problem then is formally equivalent to the strong back
scattering limit through an impurity in a TLL with an interaction
parameter $K'<1$ and a cutoff energy $\omega'_F$. This equivalence
holds not only for the Hamiltonian, but also for all the cumulants
of the current. As a quick check, the standard first-order
perturbative computation of the average current in the TLL
problem\cite{kane_fisher} yields
 Eq.(\ref{perturbativeI}). In particular, it vanishes at zero voltage, which is a consequence
of the irrelevance of tunneling. Thus any other neglected
scattering process, depending on the realistic setup, could
dominate the contribution of tunneling processes to $I$ at low
enough energy, making the prediction Eq.(\ref{perturbativeI})
non-universal.

It is much more useful to think instead of the ``dual limit'' of
weak backscattering with amplitude $v_B$, thus ${\cal
T}=(1+(v_B/hv_F)^2)^{-1}$ close to one. In the absence of coupling
to the environment, the problem is nothing but free electrons in
the presence of a potential scatterer, whose locality
 allows to use again bosonisation. A bosonic field $\Phi(x)$ determines the electronic density through
$\rho=-\partial_x\Phi/\pi$, thus the current
$j=e\partial_t\Phi/\pi$.
For pedagogical reasons, here we only present arguments at the
level of the Euclidian action. It is convenient in this limit to
integrate the bulk degrees of freedom and formulate the problem
purely in terms of the local field at the impurity
$\phi=\Phi(x=0)$\cite{kane_fisher}. If $\tau$ is the imaginary
time and $\omega_n$ are the Matsubara frequencies, one has $ \hbar
S_{el}={1\over \beta
}\sum_{\omega_n}|\omega_n||\phi(\omega_n)|^2+{v_B\over\pi
a}\int_0^{\beta} d\tau
    \cos 2\phi(\tau)$. The coupling to the impedance with a fluctuating potential drop
$e u_{env}=\hbar\partial_{\tau}\varphi$ is described by a term
$Qu_{env}$ where the transferred charge $Q$ can be identified as
$e\phi$. Thus the action acquires an additional part $\delta
S=\int d\tau \phi(\tau)(eV/\hbar-\partial_{\tau}\varphi)/\pi$
 with $V$ the
applied voltage.
%
%
Performing a partial integration over the field $\varphi$ whose
corresponding truncated action is \cite{leggett}:
$S_{env}=\sum_{|\omega_n|<\omega_R}|\omega_n||\varphi(\omega_n)|^2/(2R
e^2 \pi\beta) $ leads to a renormalization of the kinetic term,
$|\omega_n| \rightarrow {|\omega_n|}\left(1+r\right)$. There is a
formal equivalence to one impurity problem in a TLL, this time in
the weak backscattering limit and at low energy compared to
$\omega'_F=\min(\omega_F,\omega_R)$. Remarkably, one gets the same
auxiliary parameter as that found in the strong backscattering
regime, ${1\over K'}\equiv 1+r$. The auxiliary amplitude $v'_B$
will be taken as dimensionless in the following: it is
proportional to $v_B$ but depends non-trivially on the cutoffs.
The advantage of this limit is that the cosine term now defines a
relevant perturbation. Thus the predictions of the field theory
are universal as long as $v'_B$ is small enough.
 The generating Keldysh functional for $\phi$ turns out to
be identical to that in the auxiliary TLL model. Thus one can
exploit known results both for average
current\cite{apel_rice,kane_fisher} and higher cumulants defined
by \cite{saleur_weiss}:
\begin{equation}\label{defIn}
I_n=\int \left<\left< j(t_1)...j(t_n)\right>\right>_c
dt_1...dt_{n-1},
\end{equation} with $c$
indicating the connected part of the $n-th$ symmetrized current
correlator. In the following, we will introduce the differential
dimensionless cumulants:
\begin{equation}\label{defGn}G_n=\frac{h}{e^{n+1}}\frac{dI_n}{dV}.\end{equation}
 Let us first discuss
 the differential dimensionless conductance $G_1$ as inferred from lowest order perturbation
 with respect
to $v_B'$, in the limits where $k_BT/eV$ is either small or large
\cite{apel_rice,kane_fisher}:
\begin{equation}\label{G(V)WBSlow}
    G_1=\frac{h}{e^2}\frac{dI}{dV}=K'-c(K')v_B'^2
    \left(\frac{\omega}{\omega'_F}\right)^{2(-1+K')}
\end{equation}
where $\hbar\omega=\max(k_BT,eV)$, and $c(K')$ a constant
depending on $K'$. First, observe that for $v_B'=0$, one has a
purely linear regime with Ohm's law restored. The relation
$I={e^2\over h} K' V$ is obtained, which translates into
$V=(R+R_{q})I$. This is nothing but the series resistance of a
perfect point contact with resistance $R_{q}={h\over e^{2}}$ and
the resistor $R$. Second, a bare amplitude $v_B'$ is modified into
an effective larger amplitude $v_B'\omega^{-1+K'}$
 which diverges at low $\omega$ because $K'<1$, thus the above
perturbative result is valid above a voltage scale $V'_B\propto
\omega'_F v_{B}'^{1/(1-K')}$. Increasing $\omega$ up to
$\omega'_F$ drives $G_1$ to its maximum value,
$G_{max}=K'-c(K')v_B'^2$ which is still smaller than the
conductance without environment,
 $G=1-c(1)v_B'^2$. Notice that linearity can be maintained only at
 $k_BT\gg eV\gg eV'_B$, but breaks at $k_BT\ll eV$.
 On the other hand, decreasing $\omega$ below $V'_B$ increases the effective barrier
 height, and the conductance drops to zero at zero $\omega$. The low-energy
behavior of an almost transparent junction coupled
 to the environment is thus qualitatively similar to the one of a very poorly
 transmitting junction.
 In this limit, one can do perturbation with respect to a dimensionless tunnelling
 amplitude $\Gamma_T'$
 related to $v_B'$ in a a non-universal way. Thus one gets a similar result to Eq.(\ref{perturbativeI}) at $eV'_B\gg eV\gg k_BT$,
  while $eV$ has to be replaced by $k_BT$ if $eV'_B\gg k_BT\gg eV$. All these
  considerations can be made non perturbative using the
  exact solution of \cite{fendley_prl_1}.
Thus increasing the bare transparency of the conductor does not
wash out DCB but reduces its domain to $V<V'_B$.

\begin{figure}
 \vspace{0.3cm}
 \begin{center}
\epsfig{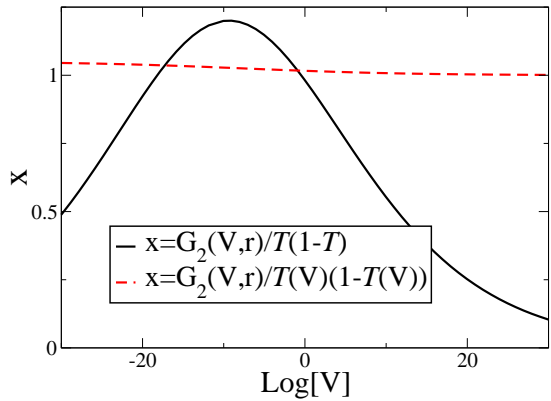}
 \caption{\label{fig3} Limit of a weak resistance, $r=0.05$: the ratio $x$ of the exact differential noise
 $G_2(V,r)$
   to the noise at $r=0$ (continuous curve), and to a "renormalized" noise given by ${\cal T}(V,r)(1-{\cal T}(V,r))$
    (dashed curve) is plotted against $\log V$ where $V$ is
    implicitly
    divided by an arbitrary voltage scale. The bare transmission is taken to be ${\cal T}=0.65$,
    while ${\cal T}(V,r)=(h/e^2)(1+r)dI/dV$.}
 \end{center}
 \end{figure}

 Motivated by
the recurrence relation between cumulants suggested in previous
works with a restriction to a weak resistance $r\ll 1$
\cite{zaikin_noise,levy_yeyati,nazarov_third_cumulant}, we now
establish
 a more general relation holding for an arbitrary $r$,
starting by a comparison of $G_1$ to the (dimensionless)
differential noise $G_2$ (Eqs.(\ref{defIn},\ref{defGn})). Let us
stick first to the two perturbative regimes so far discussed, and
to $k_BT\ll eV$, such that the noise is
poissonnian\cite{kane_fisher_noise}. More precisely $G_2\simeq 2
G_1$ for $V\ll V'_B$, while $G_2\simeq 2K' (K'-G_1)$ for $V\gg
V'_B$. Together with the expressions of $G_1$ in
Eqs.(\ref{perturbativeI}) and (\ref{G(V)WBSlow}) respectively at
low and high voltages, one can check the relation, for an {\em
arbitrary} $r$, and for $n=2$:
\begin{equation}\label{G1G2}
\frac{\partial G_{n-1}(V,r)}{\partial\log V}=-2r G_{n}(V,r).
\end{equation}
Notice that the left hand side would vanish at $r=0$, because
$G_1$ becomes voltage-independent in this limit, thus this
quantity is purely related to the presence of the environmental
resistance. This relation expresses that the dynamical Coulomb
blockade contribution to the conductance is related to the total
noise {\em in the presence of the environment}. It holds not only
at leading order in $V/V'_B$ or $V'_B/V$, but {\sl to any order}.
This truly non perturbative observation was dubbed a ``generalized
fluctuation dissipation theorem'' in \cite{fendley_prl_1}.
 In order to compare this relation to the
recent related works dealing with a small resistance $r$, we now
restrict to $K'$ close to one. At strictly vanishing $r$, $K'=1$,
and both the low and high energy series of the exact differential
conductance \cite{fendley_prl_1} can be trivially re-summed to
give the transmission probability ${\cal T}\equiv G_1= {1\over
1+v_{B}'^{2}}={\Gamma_T'^{2}\over 1+\Gamma_T'^{2}}$. To lowest
order in $r$, it is tempting to replace $G_n$ on the r.h.s of
Eq.(\ref{G1G2}) by its value at $r=0$, here $G_2(V,r)\simeq
G_2(V,0)$: doing this would suggest that the DCB contribution of a
small resistance $r$ to the current would be proportional to the
shot noise without environment as argued in
\cite{zaikin,nazarov_third_cumulant}. But one has to be careful
with the limit $r\ll 1$, as can be seen already in the two
 dual limits
 where the noise is poissonnian
(\ref{perturbativeI},\ref{G(V)WBSlow}), $G_2\sim \Gamma_T^2
V^{-2r/(1+r)}$ or $G_2\sim v_B^2 V^{2r}$; even if
 $r\ll 1$, $G_2$ can be replaced by its noninteracting value only if $V$ is not too small.
 A more quantitative comparison of $G_2(V,r)$, inferred from the exact solution of \cite{fendley_prl_1},
  to its noninteracting
 value is given by the continuous curve of Fig.(\ref{fig3}):
 the ratio $x=G_2(V,r)/G_2(r=0)$ is plotted as a function of $\log V$
  at $r=0.05$.
  Here, $G_2(r=0)={{\cal T}}(1-{{\cal T}})$ is obviously voltage-independent.
 $x$ is close to one for
  an intermediate values of voltages, but deviates from one in the limit of
 small or large voltages, a manifestation of the breakup of
 perturbation theory with respect to $r$. But the quality of this agreement depends on
 the value of the transmission coefficient, and will be discussed
 in more details
 elsewhere.


The mapping to a TLL at an arbitrary $r$ and the subsequent exact
solution can be used as well to shed some light on Ref.
\cite{nazarov_small_z} which are in the spirit of
Ref.\cite{glazman} where an effective energy-dependent
transmission coefficient ${{\cal T}}(E,r)$ is introduced, and
argued to satisfy the following RG equation in the limit of small
$r$: $\partial{{\cal T}}(E,r)/\partial \log E=-2r {{\cal
T}}(E,r)(1-{{\cal T}}(E,r))$. This formula suggests approximating
the differential noise on the r.h.s of Eq.(\ref{G1G2}), for $r\ll
1$, as $G_2(V,r)\simeq G_1(V,r)(1-G_1(V,r))$. However this is not
satisfactory at high voltages, when $G_1(V,r)\rightarrow
K'=1/(1+r)$. Rather, a better approximation is obtained by
defining ${{\cal T}}(V,r)=(1+r)G_1(V,r)$ such that $G_2(V,r)\simeq
{\cal T}(V,r)(1-{\cal T}(V,r))$, as shown through their ratio $x$
 in Fig.(\ref{fig3}) (the dashed curve). This
approximation is good up to an accuracy of $r$ in intermediate to
high voltage regimes.

 Remarkably, for an arbitrary resistance $r$, Eq.(\ref{G1G2}) can be extended to all cumulants,
 i.e. to $n>2$ where $G_n$
in Eqs(\ref{defIn},\ref{defGn}) is computed {\em with the
environmental resistance in series}, i.e. at
$K'<1$\cite{saleur_weiss}. Again, the limit $r\ll 1$ requires
care: replacing $G_n$ on the r. h. s. by its value without
environment, $G_n(V,r)\simeq G_n(r=0)$ yields the prediction of
\cite{zaikin_noise,nazarov_third_cumulant}, but with a restricted
validity domain. Rather, a better fit to \cite{nazarov_small_z} is
expected if one replaces $G_n(V,r\ll 1)$ by that expressed in the
scattering approach through the effective transmission ${{\cal
T}}(V,r)=(1+r)G_1(V,r)$, which needs to be checked.

A study of the various properties for arbitrary $K'$ and
temperature requires complex Bethe ansatz calculations, and is
identical to the examples carried out in \cite{fendley_prl_1}. The
case $K'=1/2$ is particularly simple, especially to introduce the
finite temperature. This corresponds to a crossover value, $R$
being a quantum resistance: $R=h/e^2$. While the average current
and noise have been expressed analytically, it would be
interesting to compute the higher cumulants\cite{remark_2}.

An interesting extension of these results can be done for a point
scatterer in a TLL of parameter $K$ coupled to an Ohmic
environment: the auxiliary parameter becomes ${1\over K'}={1\over
K}+{r}$, which increases the effective interactions by making $K$
smaller, thus the power law exponent is a combination of effects
of the environment and the microscopic interactions. Note however
that the role of the reservoirs, like in ordinary quantum wires,
will have to be carefully understood.

In conclusion, we have first seen in this Letter how the coupling
to an Ohmic environment induces effective repulsive e-e
interactions. While this idea is not entirely new
 \cite{neto_prl}, the setup of a well transmitting element coupled to an arbitrary
resistance provides a concrete realization, which seems very
amenable to experimental study, especially in view of the recent
progress in good transmitting atomic contacts\cite{levy_2}. It is
particularly exciting to have a potential new way of seing TLL
physics \cite{circuittll}, and the dramatic effect of a weak
backscattering barrier at low energy. Conceptually, the
relationship between TLL and dissipation is not that surprising:
starting inversely from a TLL, an electron can view the
surrounding electrons with which it interacts as an effective
electromagnetic environment\cite{nazarov_sov}. Thus a TLL can as
well be viewed as the simplest one-channel conductor coupled to a
resistor\cite{neto_prl}! Beyond its qualitative interest, the
mapping has allowed us to make contact with exact field theoretic
calculations, yielding the full counting statistics for the
current at zero temperature. We have then been able to propose a
more general link between the dynamic Coulomb blockade and the
shot noise, embodied in the exact relation (\ref{G1G2}), a
``non-equilibrium fluctuation dissipation theorem'', whose deep
origin remains somewhat mysterious, and which extends to higher
cumulants. In particular, the mapping yields the third cumulant at
arbitrary environmental resistance and transmission, opening the
perspective to include the finite temperature, especially feasible
at $r=1$.

{This problematic was motivated by D. Est\`eve, P. Joyez and C.
Urbina : we thank them for their numerous suggestions, and a
critical reading of the manuscript. I. S. would like to thank H.
Bouchiat, F. Hekking, A. Levy-Yeyati, Y. Nazarov for interesting
discussions, as well as B. Trauzettel for discussions and
providing Fig. 2.}


\begin{thebibliography}{10}
\bibitem{remark_2}
note that no exact relation similar to Eq.(\ref{G1G2}) is known
for $T\neq 0$.
\end{thebibliography}

\end{document}